\documentclass[prl,twocolumn,superscriptaddress,letterpaper]{revtex4}
\usepackage{graphicx,amssymb,bm,natbib,amsfonts,amsmath}
\usepackage{hyperref}

\begin{document}

\title{Electron-Phonon Coupling in Highly-Screened Graphene}

\author{D. A. Siegel}
\affiliation{Department of Physics, University of California,
Berkeley, CA 94720, USA}
\affiliation{Materials Sciences Division,
Lawrence Berkeley National Laboratory, Berkeley, CA 94720, USA}

\author{C. G. Hwang}
\affiliation{Materials Sciences Division,
Lawrence Berkeley National Laboratory, Berkeley, CA 94720, USA}

\author{A. V. Fedorov}
\affiliation{Advanced Light Source, Lawrence Berkeley National Laboratory, Berkeley, CA 94720, USA}

\author{A. Lanzara}
\affiliation{Department of Physics, University of California,
Berkeley, CA 94720, USA}
\affiliation{Materials Sciences Division,
Lawrence Berkeley National Laboratory, Berkeley, CA 94720, USA}

\date{\today}

%\keywords{}
\begin{abstract}
Photoemission studies of graphene have resulted in a long-standing controversy concerning the strength of the experimental electron-phonon interaction in comparison with theoretical calculations.  Using high-resolution angle-resolved photoemission spectroscopy we study graphene grown on a copper substrate, where the metallic screening of the substrate substantially reduces the electron-electron interaction, simplifying the comparison of the electron-phonon interaction between theory and experiment.  By taking the nonlinear bare bandstructure into account, we are able to show that the strength of the electron-phonon interaction does indeed agree with theoretical calculations.  In addition, we observe a significant bandgap at the Dirac point of graphene.
\end{abstract} %Reconcile this conflict by employing a new/"simple but more powerful/appropriate/" method of analysis, the result of which provides agreement for thefirst time.  Demonstrates consistency between theory and experiment.

\maketitle

The electron-electron and electron-phonon interactions are two of the fundamental interactions in many-body physics, giving rise to superconductivity, Mott-insulating behavior, and other collective phenomena.  These phenomena are often studied in association with graphene not only because graphene is a simple system featuring two carbon atoms per unit cell \cite{PRWallace} but also due to the unique potential of this material.  However, despite this apparent simplicity, there have been many difficulties in matching theoretical predictions to experimental studies of electron-phonon coupling in graphene \cite{ApparentAnisotropy}.  The nature of these discrepancies is due to the way the electron-phonon coupling constant $\lambda$ is extracted from the experimental data, and more specifically the way the bare velocity is determined.
Within the Migdal-Eliashberg regime, electron-phonon coupling results in single phonon excitations that can be treated as perturbations to the bare band dispersion and leads to a renormalization of the group velocity with respect to the electronic bare band.  The relative change of the renormalized velocity with respect to the bare velocity provides a measure of the electron-phonon coupling constant $\lambda$.  Angle-resolved photoemission spectroscopy has shown to be an invaluable probe to extract this constant since it can directly measure the single particle spectral function and hence the renormalized velocity.  However, the correct determination of $\lambda$ rests on an accurate determination of the bare velocity, which is often done by assuming a linear band approximation between the Fermi energy and high energy, a procedure that has been found to be grossly inappropriate for graphene \cite{ApparentAnisotropy}. 

The LDA band velocity is in fact known to change significantly over the relevant energy scales, which greatly affects the measured values of the electron-phonon coupling constant $\lambda$ if a linear band is assumed.  Further complicating the analysis, electronic correlations are known to renormalize the bare-band velocity in a nonlinear manner, to a degree determined by the dielectric screening of the substrate \cite{CastroNetoReview, Gonzalez0, Gonzalez1, DasSarma, Trevisanutto}.  It should also be noted that the electron-phonon coupling strength may be enhanced by interplay between electron-electron and electron-phonon interactions\cite{BaskoInterplay, BaskoRaman, Lazzeri}.  Since the experimental bare band is so difficult to determine, ARPES studies typically resort to the linear bare band approximation when determining the electron-phonon coupling strength in graphene, resulting in an over- or under-estimate of the actual electron-phonon coupling constant when extracted from the real self-energy\cite{ZhouKink,JessicaArxiv,GrapheneIr,BostwickSSCommun,VallaCaC6,Grueneis,FrictionDissipation}.

In light of these difficulties, one way to simplify the study of electron-phonon coupling in graphene might be to grow graphene on a metallic substrate, a growth technique that has recently become popular due to its relevance for technological applications\cite{CopperScience}.  On a metallic substrate, the electron-electron interaction in graphene is expected to be highly screened, which would remove velocity renormalizations due to electronic correlations and cause the bare dispersion (experimental minus electron-phonon interaction) to converge to the LDA result \cite{Hwang,Fuhrer,CastroNetoEEArxiv}.  In this highly screened limit, the curvature of the graphene LDA band structure may be taken into account when analyzing electron-phonon coupling.  Therefore the presence of a metallic substrate allows us to examine the electron-phonon interaction with an accuracy unmatched in other systems, leading to a straightforward analysis of the experimental data.
%An ideal tool for studying many-body interactions and the bare-band dispersion of graphene might be angle-resolved photoemission spectroscopy (ARPES), which would allow for a direct study of the $\pi$ bands near the Dirac point and Fermi level of graphene \cite{ZhouGap,Bostwick2007}.  ARPES spectra are a direct probe of the single-particle spectral function A(\textbf{k},$\omega$) and allow for an identification of the real and imaginary self energies Re$\Sigma$ and Im$\Sigma$, in addition to the bare-band dispersion.  
%In contrast to the cases where the effects of dielectric screening are unknown and it can be difficult to separate nonlinearities in the bare-band dispersion from self-energy effects, many of these ambiguities are expected to disappear when graphene is on a metallic substrate.  This is because the LDA band should give a good description of the bare graphene bandstructure, and any renormalizations due to electron-electron or electron-plasmon interactions should be eliminated, leading to a straightforward analysis of the experimental data.

Here we present a high-resolution ARPES study of graphene grown on a copper substrate.  Starting from a basic characterization of this system, which has never been studied before by photoemission spectroscopy, we observe sharp dispersions due to the copper substrate and graphene overlayer, including a band gap at the Dirac point of graphene.  Proceeding to examine the many-body physics in highly screened graphene, we find an overall agreement between the experimental bandstructure and the LDA band calculations.  Taking the curvature of the LDA band into account, we find close agreement between experimentally extracted electron-phonon coupling constants and theoretical calculations, providing the first real measurement of the electron-phonon coupling constant and providing closure to a long-standing debate.

Samples were grown on copper films as previously reported \cite{CopperScience}.  High-resolution ARPES data were taken at BL10.0.1 and BL12.0.1 of the Advanced Light Source at a temperature of 15$^{\circ}$K after annealing samples to 1000$^{\circ}$K, using a photon energy of 50eV.  The vacuum was better than $3\times10^{-11}$ Torr.  Potassium was deposited in situ with an SAES potassium vapor source.

\begin{figure}
\includegraphics[width=8.5 cm] {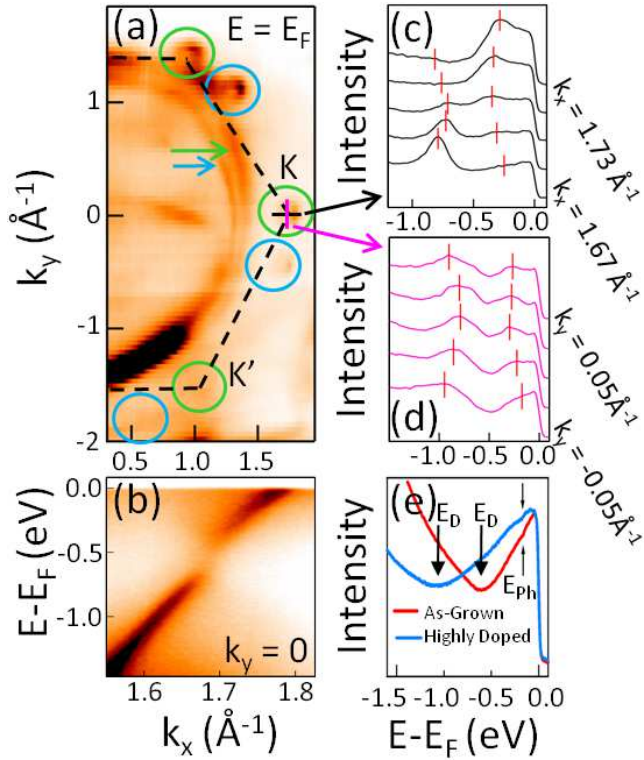}
%\label{Figure 1}
\caption{(Color online) (a) Partial map of the Fermi surface.  Arrows and circles correspond to bands from the copper substrate and graphene overlayer, respectively.  Dashed black lines correspond to the hexagonal Brillouin zone of graphene.  The K point is labelled, while the $\Gamma$-point is not shown, located at (k$_{x}$,k$_{y}$) = (0,0).  The black horizontal line through the K point illustrates the orientation of the data taken in panels (b) and (c), while the vertical purple line illustrates the orientation of the data in panel (d).  (b) ARPES dispersion and (c) EDCs taken along the $\Gamma$-K direction at k$_{y}$ = 0 \AA$^{-1}$, showing that the graphene bands are n-doped with a bandgap and intensity minimum at the Dirac point.  The presence of a bandgap creates two peaks in the EDCs at the K-point (peak positions marked in red).  (d) EDCs taken through the Dirac point along constant k$_x$.  In contrast to panel (c), where photoemission matrix elements suppress one branch of the cone, the photoemission intensity in panel (d) is symmetric and allows the presence of a bandgap to be easily seen.  (e) Angle-integrated spectra of panel (b) and of highly-doped graphene.}
\end{figure}

\begin{figure}
\includegraphics[width=8.5 cm] {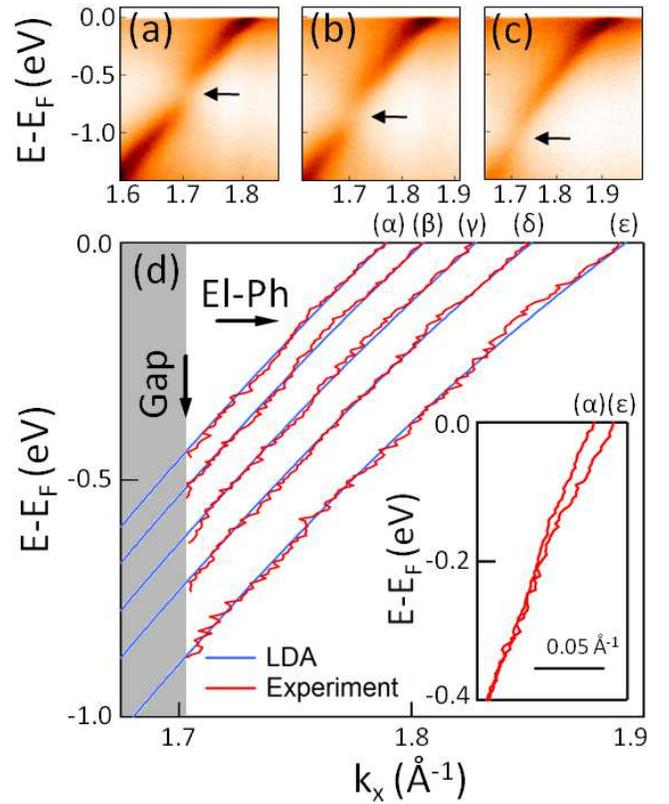}
%\label{Figure 1}
\caption{(Color online) (a-c) ARPES dispersions for several dopings, taken along the $\Gamma$-K direction.  Arrows indicate the Dirac point in each image.  Panels (a), (b), and (c) correspond to dispersions $\alpha$, $\gamma$, $\epsilon$, respectively, in panel (d).  (d) Experimental dispersions (red) for several dopings extracted from MDC peak positions, and LDA bands (blue) for the same doping along the $\Gamma$-K direction.  Greek letters $\alpha$, $\beta$, $\gamma$, $\delta$, $\epsilon$ label these dispersions in order of increasing doping.  Inset: Comparison of dispersions $\alpha$ and $\epsilon$, showing that the electron-phonon kink is stronger for the more highly doped dispersion.}
\end{figure}

\begin{figure}
\includegraphics[width=8.5 cm] {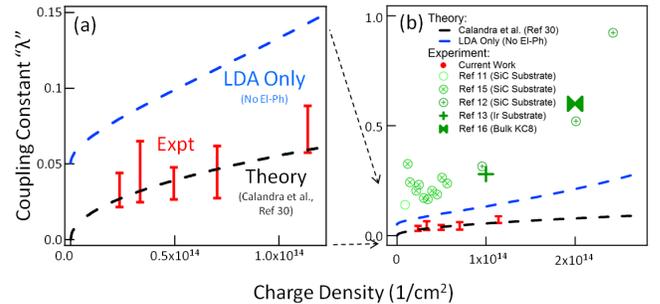}
%\label{Figure 1}
\caption{(Color online) (a) Experimental coupling constants are given as a function of electronic charge density in red.  The data agrees with theoretical calculations \cite{CalandraMauri}, shown in black.  The error generated when the linear band approximation is applied to the curved (bare) LDA band is given in blue, for comparison. (b) A zoomed-out version of panel (a), also showing results from the cited references in green (different references have different symbols depending on substrate).}
\end{figure}

Figure 1(a) shows a Fermi surface map with bands due to the copper substrate and graphene overlayers.  Two sets of copper bands and Dirac cones can be distinguished due to the presence of rotated crystallographic domains of the substrate.  The Dirac cones of graphene are visible, and are electron-doped due to their proximity to the copper substrate.  Although doping can change from sample to sample, typical values are approximately 2$\times$10$^{13}$cm$^{-2}$.  ARPES data through a single Dirac cone is shown along the $\Gamma$-K direction in figure 1(b).  In this measurement geometry, the photoemission intensity is suppressed along half of the cone\cite{Himpsel}.

The dispersion in the vicinity of the Dirac point has been the subject of some controversy in the past.  The valence and conduction bands are not collinear, possessing a region of vertical intensity between them.  Whether this is due to the presence of a bandgap in the bare dispersion or a many-body effect has been hotly debated\cite{ZhouGap,Bostwick2007}.  In the present case of graphene on a copper substrate, the dielectric screening of the highly conductive substrate rules out the possibility of electron-plasmon coupling\cite{Hwang, Polini}, and instead implies the presence of a bandgap at the Dirac point\cite{ZhouGap, EliBandgap}.
We find the separation between valence and conduction bands to be somewhat sample-dependent, with a typical bandgap of 400$\pm$50 meV, when determined from the separation between the peaks of energy distribution curves (EDCs, intensity profiles at constant momentum), as shown in figures 1c and 1d, with two peaks visible at the Dirac point momentum.  The angle-integrated intensity (figure 1e) shows a V-shaped intensity profile, with a minimum at the Dirac point energy, and increasing intensity away from the Dirac point.  Far from the Dirac point, the valence and conduction bands are not collinear, with an overall offset of 100$\pm$30 meV, suggesting an unusual band gap opening mechanism \cite{Cappelluti}. This behavior is clearly similar to results reported previously, although the size of the bandgap at the Dirac point in this sample is larger \cite{ZhouGap, EliBandgap}.  
In figure 1e, a dip can also be seen at the phonon energy, where the density of states is renormalized due to electron-phonon coupling \cite{AndreiSTS,CrommieSTM}.  In the past, calculations of electron-phonon coupling have differed from experimental results \cite{JessicaArxiv, ZhouKink, GrapheneIr, BostwickSSCommun, VallaCaC6}.  This disagreement is believed to have three sources: first, the bare band is not completely linear and may have a positive or negative second derivative depending on the magnitude of the screening of the electron-electron interaction\cite{ApparentAnisotropy,CastroNetoReview,Gonzalez0} and the direction along the brillouin zone \cite{ApparentAnisotropy}; second, the electron-electron interaction is believed to enhance the electron-phonon coupling strength \cite{BaskoInterplay}; third, the finite resolution of the experiment may lead to some error in the extracted band velocity \cite{CalandraMauri}.  The first two effects might be eliminated by the presence of a metallic substrate as this is expected to screen the electron-electron and electron-plasmon interactions, and also to eliminate the enhancement effect.  In the limit of infinite screening, metallic substrates are expected to cause the graphene dispersions to converge to LDA calculations.  This greatly simplifies the extraction of the electron-phonon self-energy.

The electron-phonon interaction is visible in the photoemission spectrum in two ways: the real part of the self energy Re$\Sigma$ modifies the band position; while the spectral width of the bands is proportional to the imaginary part of the self energy Im$\Sigma$ \cite{DamascelliReviewPaper}.  Peak positions and peak widths can be obtained by fitting lorentzian peaks to the momentum distribution curves (MDCs), the intensity at constant energy as a function of momentum.  The strength of the interaction, given by the coupling constant $\lambda$, can be extracted from either part of the self energy, although in practice the real part of the self-energy is often more reliable, since the imaginary part is more sensitive to noise and the influence of impurity broadening.  We have therefore focused on the real self-energy in our analysis.

Knowing the bare graphene band, Re$\Sigma$ is given as the difference between the experimental and bare band positions. From knowledge of Re$\Sigma$, the electron-phonon coupling constant $\lambda$ can be expressed as 
\begin{equation}
\label{eq:eq1} \lambda_k = -\left.\frac{\partial Re\Sigma_k(E)}{\partial E} \right|_{E=E_F},
\end{equation}
or equivalently,
\begin{equation}
\label{eq:eq2} \lambda_k = \frac{v_k^0(E_F)}{v_k(E_F)} - 1,
\end{equation}
where $v_k^0(E_F)$ and $v_k(E_F)$ are the bare and renormalized velocities at the Fermi level, respectively.  However, the bare band of graphene is not linear, so the method of extracting $\lambda$ according to the formula
\begin{equation}
\label{eq:eq3} \lambda_k = \frac{v_1}{v_2} - 1,
\end{equation}
(where $v_1$ and $v_2$ are the band velocities at higher and lower binding energy than the phonon, respectively) does not work, nor will any other method that assumes a linear bare band.\cite{ApparentAnisotropy}

The extracted coupling constants are compared with electron-phonon coupling calculations \cite{CalandraMauri} in figure 3.  The agreement between experiment and theory is striking, providing the first experimental support of theoretical electron-phonon calculations.  Having said this, our analysis may require a small correction.  Ab initio calculations expect a finite el-ph self-energy even at high binding energies\cite{CheolHwanElPh}.  This differs from our results, where the LDA band gives good agreement with experiment at high binding energies.  This discrepancy could correspond to a difference in the measured coupling constant of approximately 0.015$\pm$0.005, and may derive from two physical origins: 1) It is likely that the metallic substrate does not perfectly screen the electron-electron interaction in the graphene overlayer.  Since the electron-electron interaction increases the band velocity and the electron-phonon interaction decreases it, it is possible that at high binding energy both renormalizations affect the band velocity by similar amounts and essentially cancel, leaving the experimental velocity to agree with LDA.  2) It is also possible that the LDA band is not a perfect description of the bare band dispersion, due to the presence of the gap at the Dirac point.

It should also be noted that electron-phonon coupling has been studied on a metallic substrate in the past.  One study of graphene on iridium attempted to extract the electron-phonon coupling constant in a self-consistent manner, but obtained a surprisingly large value due to the approximation of a linear bare band \cite{GrapheneIr}.

To illustrate how much of a difference the linear bare band approximation makes, we have also applied the linear approximation to just the curved LDA band (which does not include electron-phonon coupling), where we take $v_k^0$ to be the slope of the line that intersects the LDA band at energies $E=E_F$ and $E=E_F$-0.4 eV using Eq. \ref{eq:eq2}.  The results, shown as the blue ``LDA'' line in figure 3, correspond to the linear bare band approximation when \textit{no} electron-phonon coupling is taking place.  Data from previous works are given in panel (b).  The error from the bare band approximation is more than twice as large as the actual electron-phonon coupling constant\cite{ApparentAnisotropy}.  On the other hand, in cases where the bare band curves in the opposite direction (``concave-up''), such as in the presence of strong electron-electron interactions\cite{CastroNetoReview, MonolayerPaper, Gonzalez0, Gonzalez1, DasSarma,Trevisanutto}, for the LDA band on the opposite side of the Dirac cone (along the M-K-Gamma direction)\cite{ApparentAnisotropy}, or for bilayer graphene\cite{FrictionDissipation}, the linear approximation underestimates the coupling constant.

In conclusion, we have shown for the first time that the magnitude of electron-phonon coupling in graphene agrees with theoretical calculations.  These results settle a long-standing controversy in the field, confirming the validity of theoretical calculations, and casting doubt on the conclusions of many experimental works.  This work is also generally applicable to future experiments that require studying electron-phonon coupling in materials with nonlinear bare bands.  We have also shown that there is a significant bandgap in graphene grown epitaxially on a copper substrate, a discovery which may pave the way for future technological applications.

\begin{acknowledgments}
We are greatly indebted to Baisong Geng and Feng Wang for providing us with the high quality graphene samples that have made this study possible.  We would also like to thank Cheol-Hwan Park for enlightening discussions.
ARPES work was supported by the Director, Office of Science, Office of Basic Energy Sciences, Materials Sciences and Engineering Division, of the U.S. Department of Energy under Contract No. DE-AC02-05CH11231.
\end{acknowledgments}

Correspondence and requests for materials should be addressed to Alanzara@lbl.gov.

\end{document}